\documentclass[11pt,preprint]{aastex}

\def\wisk#1{\ifmmode{#1}\else{$#1$}\fi}

\def\lsim   {\wisk{_<\atop^{\sim}}}
\def\gsim   {\wisk{_>\atop^{\sim}}}

\def\nwm2sr {\wisk{\rm nW/m^2/sr\ }}
\def\nw2m4sr{\wisk{\rm nW^2/m^4/sr\ }}
\shorttitle{Measuring the Mach number of the Universe.}

\received{2003 November 10}
\begin{document}
\title{Measuring the Mach number of the Universe via the Sunyaev-Zeldovich effect.}

\author{F. Atrio--Barandela
\footnote{F{\'\i }sica Te\'orica. Universidad de Salamanca. 37008 Salamanca Spain.
(atrio@usal.es)}}

\author{A. Kashlinsky \footnote{SSAI, Code 685, Goddard Space Flight Center, Greenbelt MD 20771
(kashlinsky@stars.gsfc.nasa.gov)}}

\author{J.P. M\"ucket
\footnote{Astrophysikalisches Institut Potsdam. D-14482 Potsdam.
(jpmuecket@aip.de)}}

\begin{abstract}
We introduce a new statistic to measure more accurately the 
cosmic sound speed of clusters of galaxies at different redshifts. 
This statistic is evaluated by cross-correlating cosmic microwave 
background (CMB) fluctuations caused by the Sunyaev-Zel'dovich effect 
from observed clusters of galaxies with their redshifts. When clusters are
distributed in redshift bins of narrow width, one could
measure the mean squared cluster peculiar velocity with an error 
$\sigma_{C_S^2}\lsim (300{\rm  km/s})^2$. This can be done around $z>0.3$ with clusters
of flux above 200 mJy which will be detected by PLANCK, 
coupled with high resolution microwave images to eliminate the cosmological part of the CMB fluctuations. The 
latter can be achieved with observations by the planned ALMA 
array or the NSF South Pole telescope and other surveys.
By measuring the cosmic sound speed 
and the bulk flow in, e.g., 4 spheres of $\sim 100h^{-1}$Mpc at $z=0.3$,
we could have a direct measurement of the matter density $0.21<\Omega_m<0.47$ at 95\% confidence level.
\end{abstract}


\section{Introduction}

The Mach number ${\cal M}$ characterizes the coldness of the velocity 
field. It was originally introduced by Ostriker \& Suto (1990) as a test of 
cosmological models. The cosmic Mach number
is independent of the normalization of the mass density power spectrum and 
is insensitive to redshift and bias in linear theory. It effectively measures
the {\it slope} of the power spectrum up to the scales corresponding to the size of 
the region and can be used to constrain cosmological models for structure formation (Suto \& Fujita 1990).
Strauss et al. (1993) made a direct comparison of observations and models. 
They analyzed two samples of spiral and one of elliptical galaxies and 
obtained ${\cal M}\simeq$0.5-1, with results differing with galactic type.
Nagamine et al. (2001) studied the dependence of the
cosmic Mach number on overdensity, galaxy mass and age 
and concluded that either the Local Group was in a relatively
low density region or the true mass density was $\Omega_m \sim$0.2.

Measurements of the velocity field using galaxies as distance indicators 
are affected by systematic errors (Strauss \& Willick, 1998) making 
difficult the  determination of the Mach number on scales beyond $10^4$km/s. This can be avoided  
by analyzing measurements of the Cosmic Microwave Background (CMB) as proposed in this {\it Letter}.
Hot gas in moving clusters produces redshift independent cosmic microwave background (CMB) anisotropies 
(Sunyaev \& Zel'dovich, 1972 hereafter SZ). The 
thermal component (TSZ) is usually larger than the kinetic term (KSZ), but
has a distinct spectral signature and can be removed using multifrequency
observations, allowing the determination of the cluster peculiar velocities
(Holzapfel et al. 1997; Mauskopf et al. 2000; LaRoque et al. 2002). At present,
the systematic errors are still 3 to 5 times larger than the measured velocities.  
Recently, Benson et al. (2003), used SuZIE-II measurements of 6 clusters
at $z>$0.2 to constrain the amplitude of their bulk motion,
leading to the first such determination at intermediate redshifts.

The Mach number is the ratio of the bulk flow component, which can be measured directly from the KSZ data alone
(Haehnelt \& Tegmark  1996, Kashlinsky \& Atrio-Barandela 2000,
Aghanim et al. 2001), to the cosmic sound speed. We show that the cosmic sound speed 
can be reliably measured by cross-correlating the CMB temperature field at cluster locations
{\it with their redshifts}. CMB temperature anisotropies and cluster redshifts
encode information on peculiar velocities, but have different systematic
errors. Combining the two quantities  allows to measure the average sound speed of clusters.

\section{The Cosmic Mach Number and Cosmic Sound Speed.}

The cosmic Mach number is defined to be ${\cal M}(r,R) = \frac{V_{\rm bulk}(r)}{C_{S,flow}}$.
The bulk flow component is the mean squared velocity over a region of characteristic 
scale $r$ and $C_{S,flow}$ is the sound speed in the reference frame of the mean flow.
They are: $V^2_{\rm bulk}(r)= H_0^2 f^2(\Omega_m,\Omega_\Lambda) 
\frac{1}{2\pi^2}\int_0^\infty P(k) W^2(kr) dk$, and $C_S^2 =
H_0^2 f^2(\Omega_m,\Omega_\Lambda) \frac{1}{2\pi^2}\int_0^\infty P(k) W^2(kR) dk $,
with $C^2_{S,{\rm flow}} = C_S^2 - V^2_{\rm bulk}$.
Here $f(\Omega_m,\Omega_\Lambda)\simeq \Omega_m^{0.6}+ \Omega_\Lambda
(1+\Omega_m/2)/70$ (Lahav et al. 1991), $P(k)$ is peculiar mass power spectrum and $W(x)$
is the window function of the survey.
$C_S$ depends mainly on the co-moving scales $R$ of the collapsed moving objects. 
Clusters are regions of high density and their $C_S$ may differ from the average. 
In the linear regime, Bardeen et al. (1986) showed that the rms peculiar 
velocity of peaks is smaller than that of field points, but the difference 
is less than 3\% for $R\leq 3h^{-1}$Mpc. 
Hereafter, we compute the rms peculiar velocity of clusters using linear theory
with a Gaussian kernel and $R=1.5h^{-1}$Mpc. Borgani et al. (2000) 
verified that this window and scale give a good fit to 
the measured peculiar velocities of 18 clusters from 
the ENEAR catalog of peculiar velocities of elliptical galaxies.

At present the error bars on peculiar velocities from KSZ 
($\sigma_{v_P}$) are much larger than the estimated velocities
$v_P$, preventing reliable measurement of the Mach number using CMB data alone. 
We show that measurements 
of the CMB temperature of clusters and \underline{of their redshift} can be
combined to determine ${\cal M}$ in some specific configurations.
The temperature field at a cluster location includes (Carlstrom et al. 2002):
the intrinsic CMB signal, foreground residuals, noise and TSZ ($\delta T_{Th}=
\tau k_B T_X/m_ec^2$) and KSZ ($\delta T_{Kin}= \tau v_P/c$) components. In these
expressions $\tau$ the cluster optical depth to electron scattering, 
$k_BT_X$ the intracluster electron temperature in eV,
and $v_P$ the radial component of the peculiar velocity.
All components have different amplitude depending on the beam 
of each experiment and, except the kinetic and intrinsic CMB temperature
anisotropies, all have different frequency dependence. 
The TSZ component is the dominant contribution to the CMB signal
at $l\lsim$3000 ($\lsim 4'$) (Atrio-Barandela \& M\"ucket 1999,
Molnar \& Birkinshaw 2000). Multifrequency measurements at cluster locations remove 
the temperature anisotropy down to the KSZ signal plus some residual
$\delta T = T_0 \tau (v_P/c) + r$; the latter is defined having 
zero mean and variance $ <r^2> = \sigma_{\Delta T}^2$.  
The uncertainty on the measured peculiar velocity of any individual cluster
would be $\sigma_{v_P} = c(\sigma_{\Delta T}/T_o\tau)$. 
Presently, the measurements of the KSZ (clusters with  $\tau\simeq 3-5\times 10^{-3}$), have uncertainty 
$\sigma_{v_P} \simeq 1000$km/s (Benson et al. 2003, Carlstrom et al. 2002).

The bulk flow velocity can be evaluated directly from the KSZ signal by adding the velocity of 
all clusters in the sample (Benson et al. 2003) with uncertainty:
\begin{equation}
\sigma_{V_{\rm bulk}} \simeq {\sigma_{v_P}\over \sqrt{N_{cl}}} \sim 100
\Big(\frac{\sigma_{v_P}}{1000 {\rm km/s}}\Big) \Big(\frac{N_{cl}}{100}\Big)^{-1/2} {\rm km/sec}
\label{sigma_bulk}
\end{equation}
{\it However, $\sigma_{v_P} \gg C_S$, so the error on the sound speed is 
$\sigma_{C_S^2}\propto \sigma_{v_P}^2$ and does not scale as $N_{cl}^{-1/2}$.}

The rms peculiar velocity along the line of 
sight of clusters located on a thin shell in redshift space can be reliably measured 
by cross-correlating the temperature field at cluster locations
with their measured redshifts. 
For this purpose we chose bins of width $\Delta z$ and for each cluster
we define $\delta z = z -\bar{z}$, where $\bar{z}$ is the average redshift of the
clusters in the bin. 
Cluster redshifts can be determined by photometric/spectroscopic measurements of
redshift of member galaxies or using SZ morphology (Diego et al. 2002). 
The latter would require only SZ observations but
provide an accuracy of $\sigma_z\simeq 0.04(1+z)$ for one object, while multiobject 
spectroscopic measurements of galaxies provide redshifts with 
uncertainties $\sigma_z \leq 5\times 10^{-4}$ (Yee, Ellingson \& Carlberg 1996). Thus 
redshifts of clusters in the prospective samples can be assumed to have no 
uncertainties. We also assume that  clusters are homogeneously
distributed within bins. Thus the cross-correlation gives:
\begin{equation}
\langle {\delta T\over T_o}\cdot \delta z\rangle=
\langle\tau\rangle {C^2_{S, flow}\over c^2} \pm
\langle\tau{v_{P}\over c}\cdot \delta z\rangle \pm 
\langle r\cdot \delta z\rangle .
\label{cc}
\end{equation}
Hereafter, $V_{\rm bulk}$ and $C_S$ will be the radial components
of the earlier quantities.

Peculiar velocities of clusters and their distances are not independent variables.
Clusters are chosen to be in shells of width $\Delta z$ in redshift space; this
selection criteria introduces a correlation between $d$ and $v_P$:
clusters that are most separated will have larger peculiar velocity if they 
are to be in the sample. If cluster velocities are typically $v_P\simeq C_S$,
on average one would expect the spread of clusters on real space to be:
$H_o\Delta d \simeq c\Delta z+2C_S$.  Hence:
\begin{equation}
\frac{\langle c\delta T\cdot c\delta z\rangle}{T_o\langle\tau\rangle} = 
C^2_{S, flow} \pm
{\bar\sigma_{v_P}}{c\Delta z\over\sqrt{12N_{cl}}} \pm
\langle v_p\rangle {(c\Delta z+2C_S)\over \sqrt{12}} .
\label{signal}
\end{equation}
In this expression, $\bar\sigma_{v_P}$ is the average error on the 
measured velocity of members of the
cluster sample and $\langle v_p\rangle = V_{\rm bulk}$ if the region
is smaller than the coherence length of the velocity field; otherwise
$\langle v_p\rangle = C_S/\sqrt{N_{cl}}$. 

To test eq.~(\ref{signal})
we analyzed a numerical simulation based on a P$^3$M code with 256$^3$ particles (Faltenbacher et al. 2002)
of a $\Lambda$CDM model with $\Omega_\Lambda=0.7, \sigma_8=0.87$ and Hubble constant $H_o=70$ km s$^{-1}$Mpc$^{-1}$. 
We used a ``friends-of-friends" algorithm identifying  clusters with $(1.4-23)\times 10^{14}h^{-1}$M$_\odot$. 
We randomly selected 100 shells of width $\Delta z = 0.01$ containing $\sim$100
clusters each and found that, on average, less than 10\%  of the clusters were outside
the distance interval $cH_o^{-1}$. In 80\% of the cases 
$|\langle\tau v_{P}\cdot c\delta z\rangle| \leq\langle\tau\rangle 
\langle v_p\rangle {(c\Delta z+2C_S)\over\sqrt{12}}$; for the remaining 20\%, the spread was a few percent larger.
In all cases, the second noise term in eq.~(\ref{signal}) was negligible.

The error bar of $C_S$ depends linearly on $\Delta z$.  
To reduce it we compute $C_{S, flow}$ in shells with a small spread in redshift. 
For this geometry, the dominant contribution is
\begin{equation}
 \sigma_{C_S^2,flow}=
(300 km/s)^2 \Big({\bar\sigma_{v_P}\over 1000 km/s}\Big)
\Big({\Delta z\over 0.01}\Big) 
\Big({100\over N_{cl}}\Big)^{1/2}
\label{sigma_sound}
\end{equation}
Since bulk flows in shells are small, the sound speed in the shell and the matter reference frames are
very similar: $C_{S,flow}^2\simeq C_S^2$. 
Compared with a direct estimation of $C_S^2$ from KSZ data alone, the cross correlation
with redshift represents an improvement of more than one order of magnitude.

\section{Observational Prospects.}

For small errors (Barlow, 1989): $(\sigma_{\cal M}/{\cal M})^2 = 
(\sigma_{V_{\rm bulk}}/V_{\rm bulk})^2
+ (\sigma_{C_S^2}/2 C_S^2)^2$.
In order to obtain an accurate determination, bulk flow velocities 
have to be measured in regions much smaller than the sound speed. 
We propose to compute the Mach number by estimating the sound speed in a shell
of width $\Delta z$ located at $z_{shell}$ and to measure bulk flows
in spheres of radius $R$ centered at the same redshift. 
Observations of clusters to measure their SZ amplitude are already (almost) routinely
carried out by several telescopes (e.g. Carlstrom et al. 2002) 
and exponential progress is expected with the construction of the ALMA 
array \footnote{http://www.alma.nrao.edu/} and the NSF-funded South Pole microwave 
telescope \footnote{http://astro.uchicago.edu/spt/}.
For the purpose of this project, high resolution observations
of large samples of clusters with well measured
redshifts are required. Blind surveys searching for clusters in large areas
of the sky, like the cluster catalog expected to be obtained from 
the full sky CMB observations of the upcoming 
PLANCK \footnote{http://astro.estec.esa.nl/SA-general/Projects/Planck/} 
mission or by the South Pole telescope, which will effectively observe 
$\simeq 10000$ deg$^2$, will be most relevant. 
Clusters that produce a change in flux of about 200 mJy relative
to the mean flux of the CMB will be detected by Planck. For the WMAP
$\Lambda$CDM concordance model this limit translates into 
$\simeq 9000$ clusters in $4/5$ of the sky (Diego et al. 2003). 
Cluster redshifts could be coarsely determined using SZ data (Diego et al. 2002). 
Optical follow-up of those clusters without optical counterpart
will be necessary to measure their redshifts.
The current technologies have the capacity to measure $\sim$2000 redshifts per night with 
an average uncertainty smaller than $130$ km/s\footnote{http://www.sdss.org/, http://msowww.anu.edu.au/2dFGRS/}. 
If 10 galaxies per cluster are used to define the cluster redshift, 
this procedure would require a reasonable amount of observing time.

Most clusters detected by PLANCK will be unresolved.
Those needed for computing the Mach number will have to be re-observed at 
high resolution with ALMA, where its high frequency
coverage and its small projected noise of $\sim 2\mu$K rms after
one hour of integration time for a beam of $1^\prime$, could 
be very useful to subtract foreground and CMB contribution to its minimum.
The flux limit can be related to the cluster mass (Kay et al. 2001):
$S_\nu (353 {\rm GHz}) = 75{\rm mJy} 
\Big({M_{200}\over 10^{14}h^{-1}M_\odot}\Big)^{5/3}
\Big({500 h^{-1} {\rm Mpc}\over D_A}\Big)^2 (1+z)$.
This detection threshold is quoted for the 353 {\rm GHz} Planck channel, the one with
the highest resolution and lowest noise. In this relation, $M_{200}$ corresponds
to the mass within a region where the over-density is $200\rho_c$.
Using the mass-luminosity relation (Reiprich \& B\"ohringer, 2002)
$L_{X,bol} = 1.15\times 10^{45}(M_{200}/10^{15}h^{-1}M_\odot)^{1.8}
h^{-2}$  ergs s$^{-1}$ we can estimate the luminosity of clusters detected
by Planck at any given redshift. 

\begin{figure}[ht]
\plotone{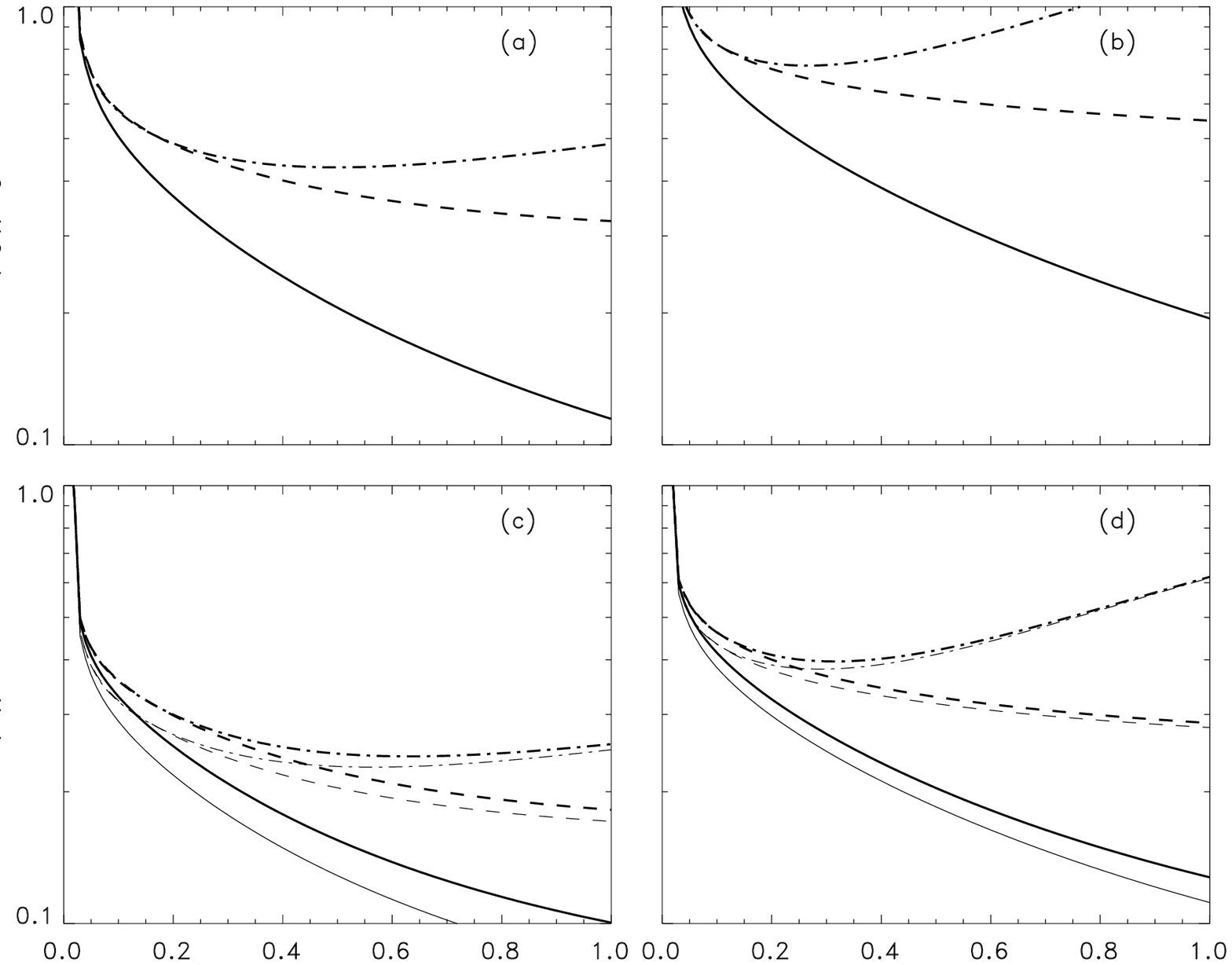}
\caption[]{
(a) Relative error of the sound speed for three different
evolutions of the luminosity function: no evolution (solid line),
dashed to $(A,B)=(-3,0)$ and dot-dashed to $(-1,-2)$ -see text-.
The detection limit is 200 mJy. (b) Same as before, but the threshold
limit is 400 mJy. (c) Relative error on the Mach number, for the
same evolution histories as before. In every pair, the upper line 
corresponds to bulk flows on a sphere of $100 h^{-1}$Mpc, and 
the lower line to a sphere of $150 h^{-1}$Mpc. The flux
limit is 200 mJy. (d) The same
as in (c) but the clusters detected have fluxes above 400 mJy.
}
\label{error}
\end{figure}

Fig.~\ref{error} shows the relative errors on $C_S^2$ and ${\cal M}$ vs $z$. The number of clusters were 
modeled with the X-ray Luminosity Function (XLF) as an evolving Schechter function:
$\phi(L,z)$=$\phi_o(1+z)^AL^{-\alpha}\exp(L/L_*)$, with $L_*$=$L_{*,0}(1+z)^B$,
where $A$, $B$ are two evolutionary parameters,
and $L_*, \alpha$ are the local XLF values. We took $L_*$=$9.3\times 
10^{44} h^{-2}$ ergs s$^{-1}$ and
$\phi_o$=$2.44\times 10^{-7}h^{-3}$Mpc$^{-3}$ (Ebeling et al. 1997).
In (a) and (b), solid lines corresponding to no redshift evolution $(A,B)$=(0,0),
dashed to (-3,0) and dot-dashed to  (-1,-2), obtained from the analysis of the
ROSAT Deep Cluster Survey and the Extended Medium Sensitivity Survey (EMSS),
respectively. At present, the evolution of
the bright end of the XLF is not well determined and the degree of
negative evolution deduced from the EMSS survey is most likely an overestimation, 
that does not extend to clusters with $L_X< 10^{44}h^{-2}$ ergs s$^{-1}$
(Rosati et al. 2002). For the 200 mJy limit (Fig. \ref{error}a)
the RDCS evolution parameters predict about 14000 clusters out to $z$=1,
and for the 400 mJy, about 6000.
Those numbers were 12000 and 4000 for the EMSS evolution parameters,
respectively, the largest difference being in the number of clusters at $z>$0.3. 
The number density of the RDCS parameters was very similar to that of Kay et al. (2002). 
We took $\bar\sigma_{v_P} $=1000 km/s, independent of $z$, even though
foreground and CMB residuals and source confusion may depend on the cluster redshift. Any $z$ dependence
can be easily accounted for since both $\sigma_{C_S^2}$ and $\sigma_{\cal M}$
vary linearly with $\bar\sigma_{v_P}$, the average error on  the velocity of 
individual clusters in the sample. 

Spheres of radius $\simeq$100$h^{-1}$Mpc
will contain very few clusters in blind all-sky surveys such as Planck 
cluster catalog, and another type of survey is required.
Romer et al. (2001) argue that an XMM serendipitous cluster
survey of about $\simeq$800 deg$^2$ 
will detect more than 8000 clusters ranging from poor to 
very rich systems. Their effective flux limit would be $\simeq 1.5\times
10^{-14}$ergs s$^{-1}$cm$^{-2}$, and all clusters above $kT_X>4$KeV
(or luminosities $L_X> 10^{44}h^{-2}$ ergs s$^{-1}$) will be detected.
Adopting the $\tau$-$L_X$ relation from Cooray (1999) 
and Kashlinsky \& Atrio-Barandela (2000),
rescaled to the bolometric band, the number cluster
density ($n_c$) verifies that $\tau (n_cR_H^3)^{1/2}\simeq$1 is
almost independent of luminosity, so that $\sim$100 clusters 
per sphere of radius $100 h^{-1}$Mpc with 
$\tau \gsim 2\times 10^{-2}$ will be detected, independent
of $z$. Also, they will have, on average, a similar error $\sigma_{v_P}$.
In Fig. \ref{error}c  and 
\ref{error}d we plot the error on the Mach number assuming that the
sound speed was computed in shells of Planck detected clusters
and the bulk velocity is computed using a top hat window on spheres of 
100 (thick) and 150 $h^{-1}$Mpc (thin lines) radii. 
Solid, dashed and dotted-dashed lines represent the same XLF evolution
models as in Figs.~\ref{error}a,b.
Fig.~\ref{error}c corresponds to Planck detecting all clusters
with flux above 200 mJy and (d) to 400 mJy.
We also considered $\bar\sigma_{v_P}$=1000 km s$^{-1}$, although now
the bulk velocity and sound speed are computed over different samples. Folding
any redshift dependence of $\bar\sigma_{v_P}$ into our results is 
straightforward, but would require detailed knowledge of the (unknown) 
cluster selection function for Planck and XMM. 

\begin{figure}[ht]
\plotone{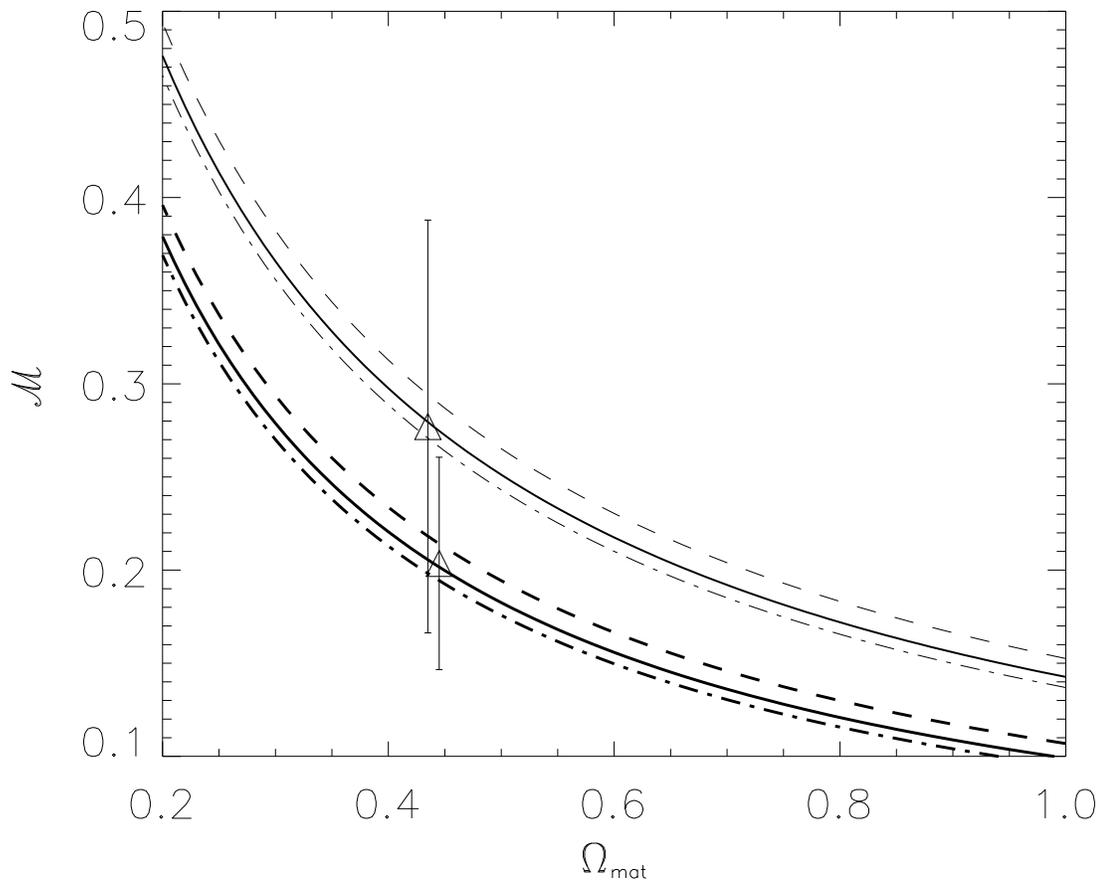}
\caption[]{Mach number estimated by measuring bulk flows on spheres of 100 (upper set)
or 150 $h^{-1}$Mpc (lower set). Dashed, solid and dot-dashed lines correspond to 
a spectral index $n=0.95, 1.0$ and $1.03$, in agreement with WMAP results.
The triangles show the prediction for the WMAP concordance model.
The 1$\sigma$ error was computed 
assuming that there were $4$ independent measurements of the
bulk flow at that redshift.
They are slightly shifted for a better display. 
}
\label{limits}
\end{figure}

To see the capacity of our method to determine cosmological parameters, we 
assume that at redshift $z\simeq $0.3 we have identified all PLANCK clusters 
with fluxes above 200 mJy in a shell of $\Delta z$=0.01, have their $z$
measured with uncertainty smaller than 150 km/s and have  them
re-observed with ALMA at $1^\prime$ resolution. For the expected number of
100 clusters, this would require $\sim$100 observing hours.  The sample is used
to compute $C_S$ at that $z$. We also assume that the clusters
identified by XMM in a sphere of radius $100h^{-1}$Mpc centered at
$z$ have data of similar quality. The Mach number would be the ratio of the
bulk velocity in the sphere to the sound speed in the shell.
Fig. \ref{limits} plots ${\cal M}$ vs matter density for flat models with cosmological
constant, $\Omega_\Lambda+\Omega_m$=1, and for three different values for the primordial spectral index
$n$ = 0.95, 1, 1.3  shown as dashed, solid and dot-dashed lines
(all other cosmological parameters were chosen from the WMAP measurements, Bennett et al. 2003, Spergel et al. 2003).
The matter power spectrum was taken as $P(k)\propto k^nT^2(k)$ with the 
analytical approximation of Sugiyama (1995) for the CDM transfer function, $T(k)$, and normalized to $\sigma_8$=0.8. 
The upper set of lines corresponds to bulk flows on spheres of 100$h^{-1}$Mpc 
and the lower set to 150 $h^{-1}$Mpc radius. 
For $\Omega_m$=0.3 the triangle shows the expected Mach number 
and its 1-$\sigma$ error bars. We assumed that the cluster XLF evolves with RDCS parameters. 
The error bar on $\cal M$ translates into a confidence interval
for $\Omega_m$. If the bulk flow is measured
for 4 independent spheres of 100$h^{-1}$Mpc radius, we can 
estimate   $0.25 (0.21) \leq\Omega_m\leq 0.37 (0.47)$ and at the 68 (95) \% confidence level.
Our method is insensitive to a running spectral index, quintessence or
tilt: variations on these cosmological parameters lead to  differences
in the Mach number much smaller than  $\sigma_{\cal M}$. In this 
respect, our measurement of the matter density would be robust.

\section{Conclusions.}

We presented a new method to measure the cosmic sound speed of clusters of galaxies. 
Assuming that all clusters with SZ flux larger than 
200 mJy will be identified by the Planck mission, and that the XRLF evolution 
in the Rosat Deep Cluster Survey is representative of the overall cluster
evolution, the relative error would be $\sigma_{C_S^2}/C_S^2 \sim 0.3$.  
If the XMM Serendipitous Cluster Survey detects all clusters above
$kT_X=4$KeV, by combining measurements of bulk flows in spheres
of different sizes with the cosmic sound speed measured in 
shells at the same redshift, we can estimate the Cosmic Mach number with 
relative accuracy $\sigma_{\cal M}/{\cal M}\geq 0.2$.
This accuracy can be achieved with high resolution
microwave images in order to reduce the intrinsic CMB anisotropy with respect
to the Kinematic SZ signal, and is within the reach of the currently planned
South Pole and ALMA telescopes.
 The Cosmic Mach number we can determine by  our method is directly related to 
the matter density, and it can determine $0.21\leq\Omega_m\leq 0.47$
at the 95\% confidence level.
This result, being independent of other mass density measurements,
would provide an important self consistency check.

FAB and JPM thank the financial support of the Spanish-German Acciones Integradas
program HA2002-0084. FAB also thanks the Junta de Castilla y Le\'on research project SA002/03
and Ministerio de Ciencia y Tecnolog{\'\i}a projects BFM2000-1322 and
AYA2000-2465-E.

\clearpage

\clearpage

\end{document}